\documentclass{article}
\begin{document}

\begin{center}
\centerline{\large \bf Physical origin of "laser induced" molecule alignment 
effect.}
\end{center}

\vspace{3 pt}
\centerline{\sl V.A.Kuz'menko}
\vspace{5 pt}
\centerline{\small \it Troitsk Institute for Innovation and Fusion 
Research,}
\centerline{\small \it Troitsk, Moscow region, 142190, Russian 
Federation.}
\vspace{5 pt}
\begin{abstract}

	The alternative to dynamic alignment explanation of experimental 
results on spatial-asymmetric dissociation of molecules in a laser field 
is proposed. The concept of geometrical alignment is sufficient for 
explanation of these results. In this case the spatial anisotropy of 
interaction of molecules with laser radiation is transferred from one 
field to another through the ordinary mechanism of nonlinear optical 
interactions. Thus laser radiation does not create alignment, but only 
registers it. A physical basis of such nonlinear processes is inequality 
of forward and reversed optical transitions that corresponds to a concept 
of time invariance violation in electromagnetic interactions. Directions 
of the further researches in the field of alignment spectroscopy are discussed.

\vspace{5 pt}
\end{abstract}

\vspace{12 pt}

Last years the impressing successes are achieved in the field of control of 
atoms motion with the help of laser radiation [1]. Laser radiation allows to 
keep the atoms long time and to cool its up to ultra low temperature. 

	For molecules the situation is quit different. In the field of 
keeping and cooling of molecules by laser radiation only minimal successes 
are achieved [2]. Here the effects, connected to orientation of molecules in 
space are usually studied. These effects are divided into two types: passive 
(geometrical) and active (dynamic) alignment [3]. The concept of geometrical 
alignment is connected with the generally recognized opinion, that the 
probability of optical transition depends on orientation of the molecule in 
space relative to the direction and polarization of the laser beam [4]. How 
strong is this dependence? What is it shape? Such information is absent for 
present day. However, it has a fundamental meaning. At the same time the 
opportunity to obtain such information exists and will be discussed below.

	The dynamic alignment effect usually is connected with the assumption 
that intense laser field is capable to create the certain orientation of 
molecules in space. It is supposed, that the dipole moment induced in a 
polarizable molecule by the laser field can hinder free rotation of 
molecules. Molecules appear in the so-called pendular states, in which the 
molecular axis librates about the electric field vector. When the laser 
radiation is switched off, the molecules continue free rotation [5, 6]. 
Such assumption seems fantastic. The situation is interesting, because of 
any reliable experimental proofs of existence of the dynamic alignment 
effect are absent. At the same time this assumption is widely used for an 
explanation of experimental results. In turn, these experimental results are 
used as the proofs of the existence of the effect of dynamic alignment.

	Some doubts in accuracy of such proof arose earlier [7]. The purpose 
of the present article is to add the big piece of such doubts and to show, 
that for explanation of experimental results there are enough conception 
about the geometrical alignment which anisotropy can be transferred from one 
laser field to another by ordinary nonlinear interactions. 

	We shall consider in more detail typical experiment on dynamic 
alignment of iodine molecules [8]. Cooled to low temperature in jet the 
iodine molecules interact with the powerful nonresonant laser pulse with 
linear polarization. Simultaneously with it or later the molecules are 
dissociated by the second powerful laser pulse. Then the fragments of 
molecules are ionized and their angular distributions are studied. Thus 
without the first laser field the dissociation of molecules is isotropic. 
However, when the first laser pulse is present, the strong spatial anisotropy 
of molecule dissociation is observed. The effect is explained as a result of 
the dynamic alignment of molecules and, in turn, it serves as the proof of 
existence of dynamic alignment.

	The alternative explanation assumes the lack of any dynamic 
alignment. There exist only the dependence of optical transition probability 
from orientation of molecules in space (geometric alignment effect) and 
nonlinear optical phenomena. Laser radiation does not create orientation of 
molecules, but only registers it. The situation can be discussed in a popular 
now terms of optical information storage [9-11]. Radiation of the first laser 
writes down the information and radiation of the second laser reads out this 
information. Reading of the information occurs only at a certain orientation 
of the molecules in space, which is set by a direction of the first laser 
beam. Thus the effect of geometrical alignment is transferred from one laser 
field to another.

	The physical mechanism of process can be described as follows. Under 
action of the first laser field the Raman transitions occur between various 
rotational sublevels of ground state of iodine molecules (Fig.1). Thus the 
laser radiation interacts most effectively only with the molecules, which 
have appropriate orientation in space. So, the effect of geometric alignment 
takes place. The essence of the explanation takes into consideration, that 
the spectroscopic properties of molecules, which participated in Raman 
transition, differ from those for molecules, which did not interact with 
laser radiation (usually similar problems are discussed within the framework 
of the "coherency" concept). The molecules have property of memory about the 
initial state and aspiration to return in this condition.

	To get precisely in the initial state the molecule should have 
certain orientation in space concerning the direction of the second laser 
beam. Under action of the second laser field the reversed in the initial 
state Raman transition occurs. The experiments show, that the cross-section 
of the reversed in the initial state optical transition can considerably 
exceed cross-section of the forward optical transition [12]. Therefore 
radiation of the second laser field interacts mainly with the molecules, 
which participated in forward optical transitions. If the quantum of the 
second laser field is great enough, or a multiphoton process takes place, 
the transition into the initial state is accompanied by dissociation of 
grater or smaller part of molecules. Thus the effect of geometric alignment 
is transferred from one laser field to another as a result of ordinary 
process of four-photon mixing.

	Representation about inequality of forward and reversed optical 
transitions lays in a basis of the given alternative explanation of "induced" 
anisotropy of dissociation of molecules. The same representation underlies a 
physical explanation of the majority nonlinear optical phenomena [12,13]. 
This representation corresponds to a principle of time invariance violation 
in electromagnetic interactions. Such conclusion is very strong, but it has 
a physical basis [14] and the direct experimental proof [12]. Recently the 
new experimental data about T-invariance violation in optics are received 
[15,16]. Authors believe that they find unambiguous experimental evidence of 
a time non-reversal interaction between light and metallic planar chiral 
nanostructures. 

	One of the most important problems of experimental researches in 
optics is independent measurement parameters of forward and reversed optical 
transitions. It is an uneasy task [12]. The certain information on the given 
problem can be received in the field of discussed experiments on alignment 
spectroscopy. We keep in mind a more detailed studying the transition of 
process from nonadiabatic to adiabatic mode. Under adiabatic regime the case 
is understood, when the duration of the front and back edges of the first 
laser pulse exceeds the duration of the rotational period of molecules. 
In this case all molecules have an opportunity to interact effectively 
with laser radiation. The alignment effect is observed during action of the 
first laser pulse, but is absent after its ending. The physical essence of 
the phenomenon, probably, can be explained as follow: the front edge of a 
laser pulse carries out the forward Raman transitions and the low intensive 
back edge of a laser pulse carries out mainly the reversed Raman transitions 
in the initial state. As a result the memory of molecules is deleted and 
the "coherency" disappears (in nonlinear optics the concept of inequality 
of forward and reversed transitions is, obviously, a physical equivalent, 
base of the concept of "coherency"). 

	When the short laser pulse is used (much less than the duration of 
the molecular rotational period), the nonadiabatic mode is realized. In this 
case only small part of molecules, which have proper orientation in space, 
interacts effectively with laser radiation. These molecules can not return in 
the initial state and the alignment effect is observed for a long time as 
short periodic revivals [17,18]. So, the detailed studying of efficiency of 
the back edge of the pulse on suppression of alignment effect can give certain 
information about ratio of cross-sections of forward and reversed optical 
transitions.

	The major direction of alignment spectroscopy should become 
experimental studying of dependence of probability of optical transition 
from orientation of molecules in space relative to the direction of a laser 
beam. The basic way for getting of such information will consist in studying 
the consecutive optical transitions in crossed laser beams. The simplest 
example of such transitions is a Raman transition under action of two laser 
fields with different frequency. In work [19,20] the directions of two laser 
beams coincided and the authors observed, that the character of spectrum was 
changed when the polarization direction of second laser field was rotated. 
The effect was explained as the result of dynamic alignment and creation of 
pendular states. Obviously, it is wrong interpretation. We have two 
consecutive optical transitions. On the first transition the radiation of 
the first laser selects molecules with the certain orientation. That is the 
geometrical alignment takes place. Radiation of the second laser induces 
transition in these already selected and oriented molecules. So, the 
character of this transition should depend on the direction and polarization 
of radiation of the second laser beam. 

	The next step should be experimental studying of optical transitions 
in crossed laser beams. Dependence of efficiency of Raman transition on the 
angle between the laser beams will give the information about the dependence 
of probability of optical transition from orientation of molecules in space 
relative to a direction of the laser beam. It is possible to use various 
optical schemes on the basis of a four-photon mixing process. It is possible 
to try to receive the information separately about forward and separately 
about reversed optical transitions. It is an extensive work for many years.

	In conclusion, the alternative explanation of the molecule dynamic 
alignment effect is proposed. It takes into account only the geometric 
alignment effect and nonlinear character of a molecule-field interaction. 
A basis of nonlinear character of such interactions is inequality of forward 
and reversed optical transitions that corresponds to a principle of time 
invariance violation in electromagnetic interactions. Directions of the 
further researches in the field of alignment spectroscopy are discussed.

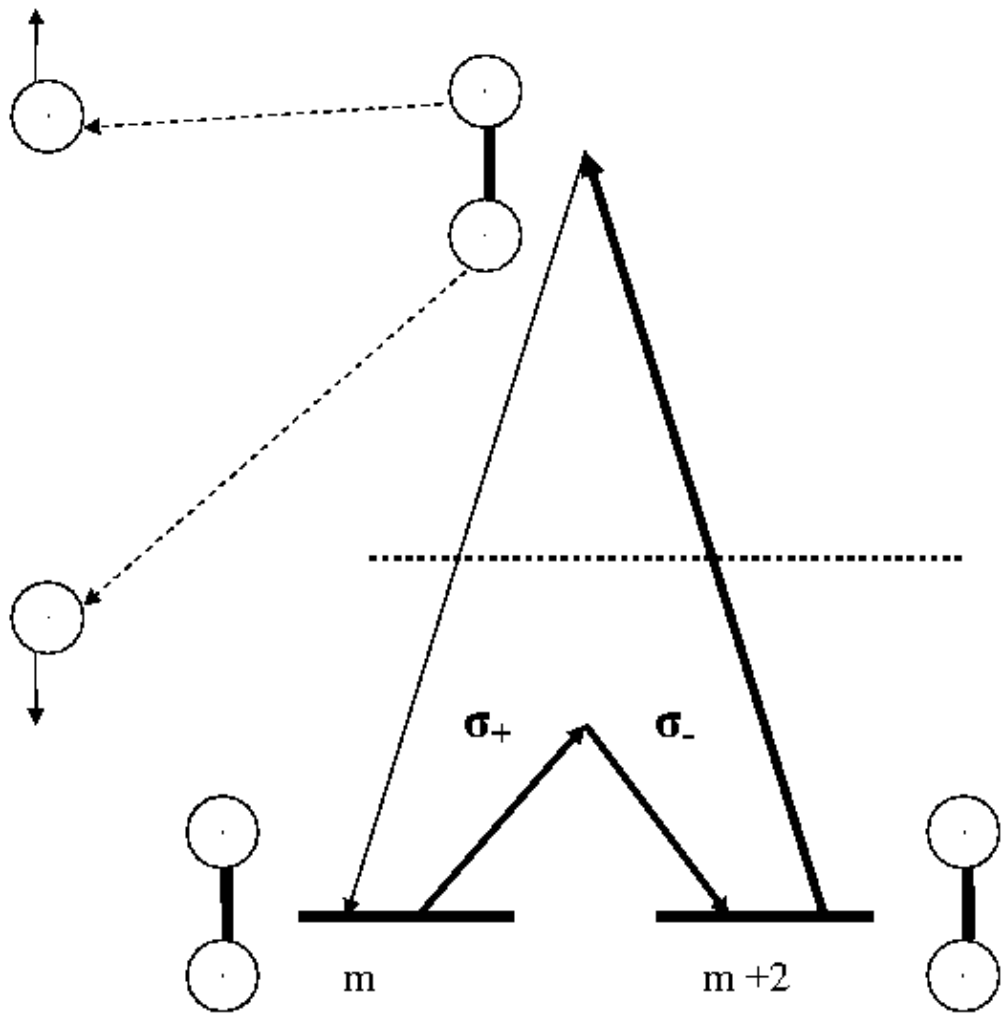
Fig.1  Four-photon mixing scheme for alternative explanation of molecule
 alignment experiments.


\vspace{5 pt}

\begin{thebibliography}{99}
\bibitem{1} H.J.Metcalf and P.van der Straten, \emph {Laser cooling and 
trapping}, Springer-Verlag, Berlin, (1999). 
\bibitem{2} H.Sakai, A.Tarasevitch, J.Danilov, H.Stapelfeldt, R.W.Yip, 
C.Ellert, E.Constant, and P.B.Corkum, Phys.Rev.A {\bf 57}, 2794 (1998). 
\bibitem{3} S.Banerjee, G.R.Kuman, and D.Mathur,  Phys.Rev.A {\bf 60}, 
R3369 (1999).
\bibitem{4} R.C.Estler and R.N.Zare, J.Am.Chem.Soc. {\bf 100}, 1323 (1978).
\bibitem{5} B.Friedrich and D.Herschbach, Phys.Rev.Lett. {\bf 74}, 4623 (1995).
\bibitem{6} H.Stapelfeldt and T.Seideman, Rev.Mod.Phys. {\bf 75}, 543 (2003).
\bibitem{7} Ch.Ellert and P.B.Corkum, Phys.Rev.A {\bf 59}, R3170 (1999).
\bibitem{8} J.J.Larsen, H.Sakai, C.P.Safvan, I.Wendt-Larsen, and 
H.Stapelfeldt, J.Chem.Phys. {\bf 111}, 7774 (1999).
\bibitem{9} A.V.Matsko, Y.V.Rostovtsev, O.Kocharovskaya, A.S.Zibrov, and 
M.O.Scully, Phys.Rev.A, {\bf 64}, 043809 (2001).
\bibitem{10}D.F.Phillips, A.Fleischhauer, A.Mair, R.L.Walsworth, and 
M.D.Lukin, Phys.Rev.Lett., {\bf 86}, 783 (2001).
\bibitem{11} H.Gao, M.Rosenberry, and H.Batelaan, Phys.Rev.A, {\bf 67}, 
053807 (2003).
\bibitem{12} V.A.Kuz'menko, E-print, physics/0306148.
\bibitem{13} M.Xiaochun, E-print, physics/0308037.
\bibitem{14} J.S.M.Ginges and V.V.Flambaum, E-print, physics/0309054, p.49.
\bibitem{15} A.Papakostas, A.Potts, D.M.Bagnall, S.L.Prosvirnin, H.J.Coles, 
and N.I.Zheludev, Phys.Rev.Lett., {\bf 90}, 107404 (2003).
\bibitem{16} A.S.Schwanecke, A.Krasavin, D.M.Bagnall, A.Potts, A.V.Zayats, 
and N.I.Zheludev, E-print, cond-mat/0307056.
\bibitem{17} E.Peronne, M.D.Poulsen, Ch.Z.Bisgaard, H.Stapelfeldt, and 
T.Seideman, Phys.Rev.Lett., {\bf 91}, 043003 (2003).
\bibitem{18} P.W.Dooley, I.V.Litvinyuk, K.F.Lee, D.M.Rayner, M.Spanner, 
D.M.Villeneuve, and P.B.Corkum, Phys.Rev.A, {\bf 68}, 023406 (2003).
\bibitem{19} W.Kim and P.M.Felker, J.Chem.Phys., {\bf 104}, 1147 (1996).
\bibitem{20} W.Kim and P.M.Felker, J.Chem.Phys., {\bf 107}, 2193 (1997).
\end{thebibliography}
\end{document}